\documentclass[conference,letterpaper]{IEEEtran}
\IEEEoverridecommandlockouts

\usepackage[utf8]{inputenc} 
\usepackage[T1]{fontenc}
\usepackage{url}
\usepackage{ifthen}
\usepackage{cite}
\usepackage[cmex10]{amsmath} 

\usepackage{textcase}
\usepackage[tablename=Table]{caption}
\usepackage{blindtext}
\usepackage{floatrow}
\usepackage{soul}
\usepackage{changes}
\usepackage{gensymb}
\usepackage{cite}
\usepackage{amsmath,amssymb,amsfonts}
\usepackage{textcomp}
\usepackage{graphicx}
\usepackage{amssymb}
\usepackage{amsfonts}
\usepackage{amsmath}
\usepackage{epsfig}
\usepackage{color}
\usepackage{fancybox}
\usepackage{textcomp}
\usepackage{multirow}
\usepackage{setspa ce}
\usepackage{psfrag}
\usepackage{booktabs}
\usepackage{float}
\usepackage[caption = false]{subfig}
\usepackage{algorithm}
\usepackage{algpseudocode}
\usepackage{mathtools, nccmath, bigints, amsfonts}

\usepackage{array}
\newcolumntype{L}{>{\centering\arraybackslash}m{3cm}}

\usepackage{mathrsfs}

\newfloatcommand{capbtabbox}{table}[][0.4\textwidth]

\newtheorem{Remark}{Remark}

    \def\Complex{{\rm\rule[.23ex]{.03em}{1.1ex}\kern-.3em{C}}}

    \newcommand{\be}{\begin{equation}} \newcommand{\ee}{\end{equation}}
    \newcommand{\bea}{\begin{eqnarray}} \newcommand{\eea}{\end{eqnarray}}
    \newcommand{\benum}{\begin{enumerate}} \newcommand{\eenum}{\end{enumerate}}

    \newcommand{\qa}{{\bf a}}
        \newcommand{\qb}{{\bf b}}
        \newcommand{\qc}{{\bf c}}

        \newcommand{\qg}{{\bf g}}
        \newcommand{\qh}{{\bf h}}

        \newcommand{\qm}{{\bf m}}
        \newcommand{\qn}{{\bf n}}

        \newcommand{\qq}{{\bf q}}

        \newcommand{\qu}{{\bf u}}

        \newcommand{\qx}{{\bf x}}
        \newcommand{\qy}{{\bf y}}

        \newcommand{\qA}{{\bf A}}
        \newcommand{\qB}{{\bf B}}

        \newcommand{\qH}{{\bf H}}
        \newcommand{\qI}{{\bf I}}

        \newcommand{\qV}{{\bf V}}
        \newcommand{\qW}{{\bf W}}

        \newcommand{\qzero}{{\bf 0}}

        \newcommand{\qSigma}{{\boldsymbol \Sigma}}

        \newcommand{\qlambda}{{\boldsymbol \lambda}}
        \newcommand{\qgamma}{{\boldsymbol \gamma}}

        \newcommand{\qmu}{{\boldsymbol \mu}}

        \newcommand{\calN}{{\mathcal N}}

        \newcommand{\Ex}{{\sf E}}

\def\BibTeX{{\rm B\kern-.05em{\sc i\kern-.025em b}\kern-.08em
    T\kern-.1667em\lower.7ex\hbox{E}\kern-.125emX}}

\begin{document}

\title{{Graph Neural Network Aided Expectation Propagation Detector for MU-MIMO Systems}}

\author{\IEEEauthorblockN{Alva Kosasih\IEEEauthorrefmark{1}, Vincent Onasis\IEEEauthorrefmark{1}, Wibowo Hardjawana\IEEEauthorrefmark{1}, Vera Miloslavskaya\IEEEauthorrefmark{1}, \\ Victor Andrean\IEEEauthorrefmark{2}, Jenq-Shiou Leu\IEEEauthorrefmark{2}, Branka Vucetic\IEEEauthorrefmark{1} }\\
\IEEEauthorblockA{\IEEEauthorrefmark{1}Centre of Excellence in Telecommunications, The University of Sydney, Sydney, Australia. \\ \IEEEauthorrefmark{2}Mobilizing Information Technology Lab., National Taiwan University of Science and Technology, Taipei, Taiwan. \\ 
Email:\{alva.kosasih,wibowo.hardjawana,vera.miloslavskaya,branka.vucetic\}@sydney.edu.au }
 vona0880@uni.sydney.edu.au,andreanvictor6374@gmail.com,jsleu@mail.ntust.edu.tw.}


\maketitle

\begin{abstract}
Multiuser massive multiple-input multiple-output  (MU-MIMO) systems can be used to meet high throughput requirements of 5G and beyond networks. In an uplink MU-MIMO system, a base station is serving a large number of users, leading to a strong multi-user interference (MUI). Designing a high performance detector in the presence of a strong MUI is a challenging problem. This work proposes a novel  detector based on the concepts of expectation propagation (EP) and graph neural network, referred to as the GEPNet detector, addressing the limitation of the independent Gaussian approximation in EP. The simulation results show that the proposed GEPNet detector significantly outperforms the state-of-the-art MU-MIMO detectors in strong MUI scenarios with equal number of transmit and receive antennas.
\end{abstract}


\section{Introduction}

Multiuser massive multiple-input multiple-output (MU-MIMO) technique is one of the key
technologies to enable a high throughput in 5G and beyond networks \cite{Borges2021}. The usage of multiple transmit and receive antennas ensures a high spectral efficiency \cite{1998Foschini_WCommun_MultiuserMIMO}, and therefore a high throughput. 
One of the challenging problems in uplink MU-MIMO systems 
is to design a practical base station detector that can achieve a high reliability performance in the presence of a strong multi-user interference (MUI). The MUI is caused by multiple user antennas simultaneously sending information to multiple base station antennas. 
The state-of-the-art practical MU-MIMO  detectors can be classified as classical and neural network (NN)-based detectors.

The classical detectors \cite{2009Donoho_ProcSci_AMP,2017SRangan_ISIT_VAMP,Ma-17ACCESS,Jespedes-TCOM14} use Gaussian distributions to approximate the posterior probability of the transmitted symbol estimates conditioned on the received signal. They were shown to achieve a near maximum likelihood (ML) performance \cite{ML} only when the number of receive antennas is much higher than the number of transmit antennas (users).
The approximate message passing (AMP) detector \cite{2009Donoho_ProcSci_AMP} 
performs poorly in the case of ill-conditioned channel matrices. 
The problem of ill-conditioned channel matrices has been partially resolved by the orthogonal AMP (OAMP) detector  \cite{Ma-17ACCESS} by integrating the linear minimum mean square error (MMSE) filtering.
The expectation propagation (EP) detectors  \cite{2017SRangan_ISIT_VAMP,Jespedes-TCOM14} outperform the OAMP detector by introducing regularization parameters in the MMSE filter that are adjusted iteratively according to the channel matrix and MUI level. 
However, there is still a significant performance gap between the EP and ML detectors when the number of base station receive antennas is equal to the number of user transmit antennas, referred to as a high MUI scenario. 

The NN-based detectors have been proposed in   \cite{Corlay_2018,2019_Samuel_TSP_Detnet,AScotti_GNN_2020,2018HHE_Globecom_OAMPNet,2021_KPratik_TSP_REMimo} to  address the performance limitation of the mentioned classical detectors in the case of ill-conditioned channel matrices and/or high MUI. This is done by unfolding their iterations into NN layers and optimizing their parameters. 
The OAMPNet detector \cite{2018HHE_Globecom_OAMPNet} combines the OAMP and NN that has a small number of trainable parameters to deal with the ill-conditioned channel matrices.
This results in a significant performance improvement compared to the conventional OAMP detector.
A high performance recurrent equivariant (RE)-MIMO detector  was proposed in \cite{2021_KPratik_TSP_REMimo}. The RE-MIMO detector unfolds the  AMP detector and integrates it with a transformer self-attention network to cancel the high MUI and ensure equivariance under permutations of the user transmit antennas. The addition of the transformer based MUI canceller results in a significant performance improvement compared to the conventional AMP detector.
Nevertheless, a significant performance gap remains when comparing the performance of the NN-based and ML detectors 
in  a high MUI scenario \cite{2021_KPratik_TSP_REMimo,2018HHE_Globecom_OAMPNet}. Our analysis shows that the reason is the inaccuracy of the Gaussian approximation. To the best of the authors' knowledge, none of the state-of-the-art detectors address this issue.

In this paper, we propose a  novel unfolded  NN-based detector for high MUI scenarios, referred to as graph EP network (GEPNet) detector. The proposed detector integrates the EP \cite{Jespedes-TCOM14} and graph neural network (GNN) \cite{AScotti_GNN_2020} as follows.
The EP can be divided into three modules: (1) an observation module, calculating the likelihood function of the transmitted symbols based on the received signal; (2) a Gaussian approximation module, approximating the posterior probability distribution of each transmitted symbol estimate using a Gaussian distribution; and (3) an estimation module, calculating the transmitted symbol estimates. The second module assumes that the joint posterior probability distribution of the transmitted symbol estimates is approximated by the product of $K$ independent Gaussian  distributions, where $K$ is the number of users. 
As a consequence, the EP loses some MUI information. 
In MU-MIMO systems with high MUI, this approximation 
is therefore inaccurate and induces a severe performance degradation. Instead of using this approximation, the proposed detector uses the GNN to produce the posterior probability distribution parameterized according to a Markov random field (MRF). Specifically, we adopt a factor graph representation \cite{Forney}. The MUI between each pair of users is characterized by a pair potential. Thus, the GNN  captures  the  MUI information  using the MRF.  The main contributions of this paper are unfolding EP into NN layers and integrating it with the GNN to address the limitation of the independent Gaussian approximation in EP. 
This contribution results in the first offline NN-based detector.
In contrast to all existing classical \cite{2009Donoho_ProcSci_AMP,2017SRangan_ISIT_VAMP,Ma-17ACCESS,Jespedes-TCOM14}   and NN-based \cite{2019_Samuel_TSP_Detnet,2018HHE_Globecom_OAMPNet,AScotti_GNN_2020,2021_KPratik_TSP_REMimo} detectors, the proposed GEPNet detector can achieve a high detection performance in a high MUI scenario 
and significantly improves the EP performance. To the best of the authors' knowledge, 
the GEPNet is the first detector outperforming the EP by replacing the independent  Gaussian  approximation.
The simulation results show that the GEPNet detector outperforms the EP, OAMPNet, and RE-MIMO  detectors by more than $4$ dB at the SER of $10^{-4}$ for $64\times 64$ MU-MIMO configuration. 

{\bf Notations}: $\qI_n$ denotes an identity matrix of size $n$.  For any matrix $\mathbf{A}$, the notations $\mathbf{A}^{T}$  and $\mathbf{A}^{\dagger}$ stand for transpose and pseudo-inverse of $\mathbf{A}$, respectively.  $\|\qq\|$ denotes the Frobenius norm of vector $\qq$.   $q^*$ denotes the complex conjugate of a complex number $q$. Let $\qx = [x_1, \cdots, x_K]^T$ and $\qc = [c_1, \cdots, c_K]^T$.
 ${\Ex}[\qx]$ is the mean of random vector $\qx$, and ${\mathrm{Var} }[\qx] = {\Ex}\big[\left(\qx-{\Ex}[\qx]\right)^2\big]$ is its variance.   $\calN(x_k: c_k,v_k)$ represents a single variate Gaussian distribution  for a  random variable $x_k$ with mean $c_k$ and variance $v_k$.  

\begin{figure}
\centering
{\includegraphics[scale=0.34]{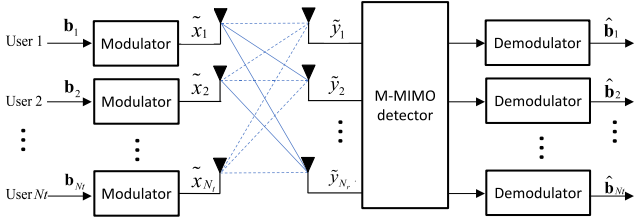}}
\caption{The MU-MIMO system}
\label{up_MU-MIMO}
\end{figure}

\section{System Model}

\begin{figure*}
\centering
\subfloat[The iterations in the GEPNet detector]
{\includegraphics[scale=0.40]{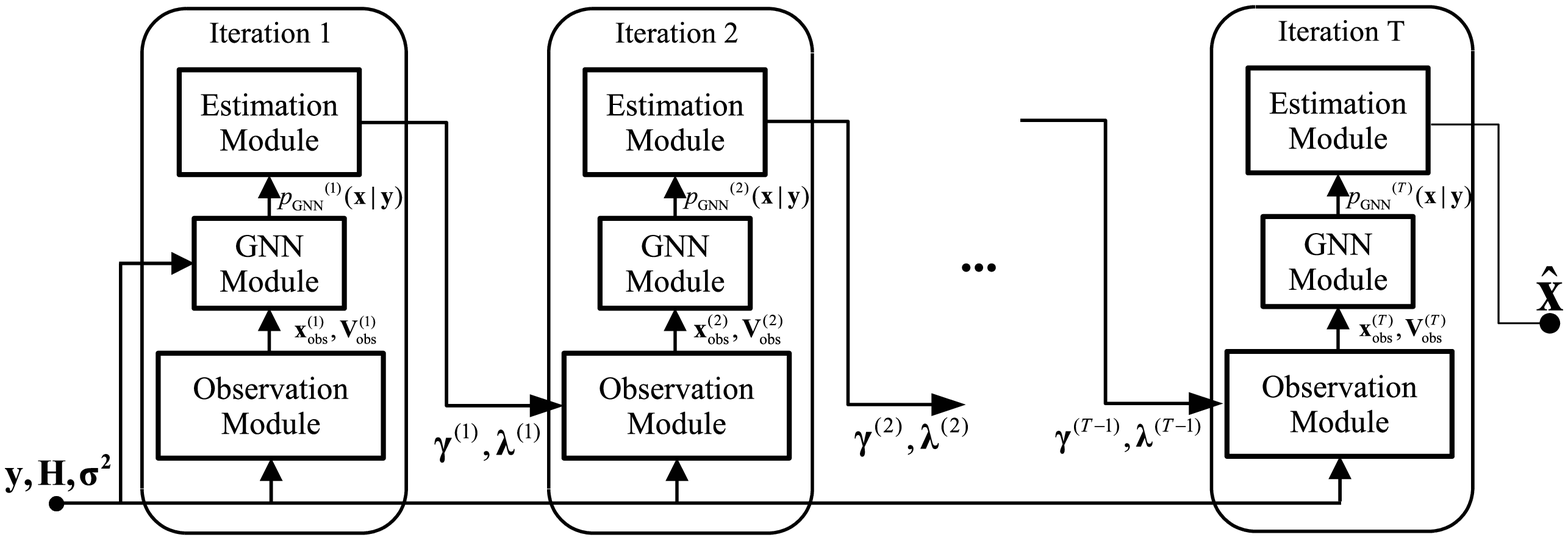}}\hfill
\centering
\subfloat[The MPs in the $t$-th iteration ]
{\includegraphics[scale=0.40]{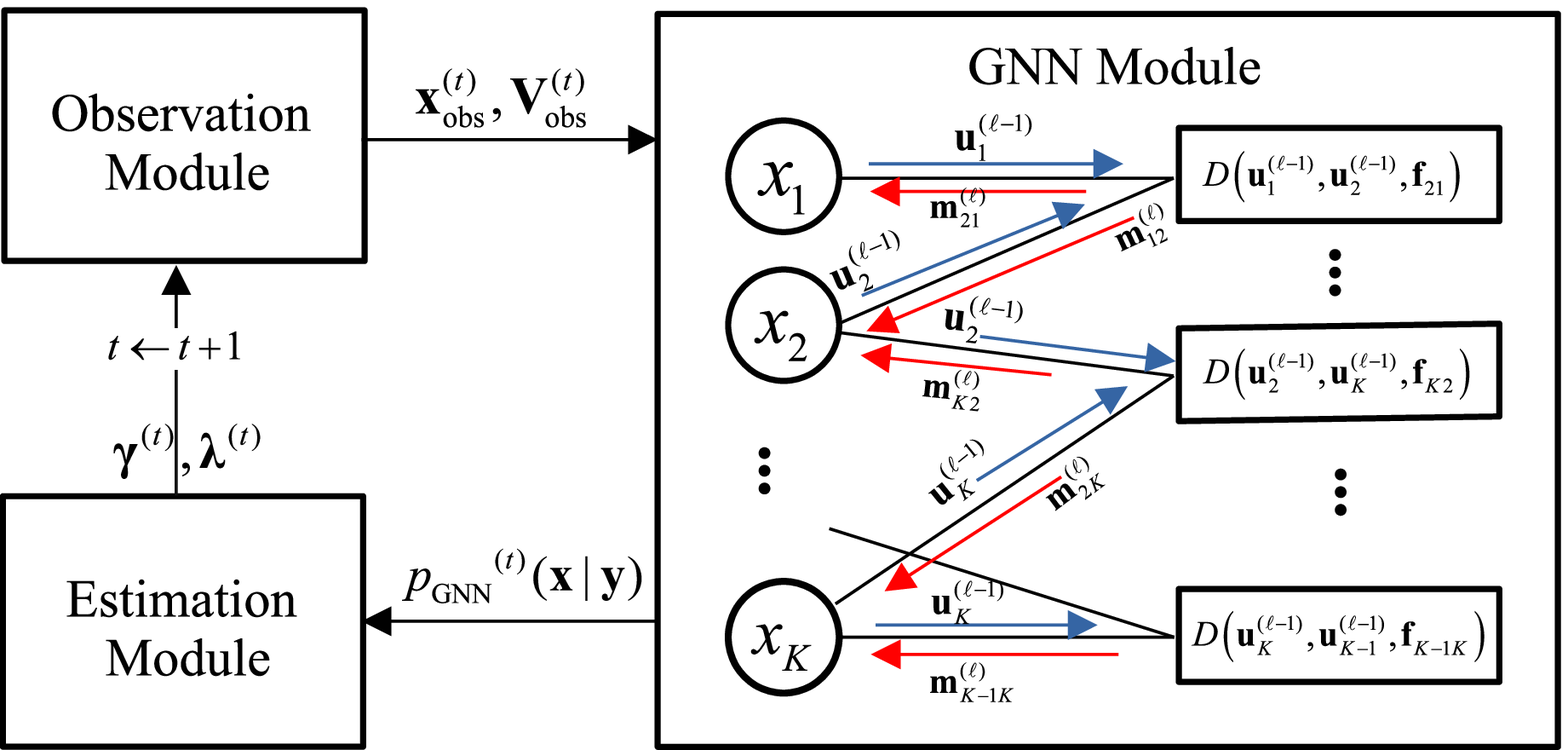}}
\caption{The GEPNet detector model}
\label{F2}
\end{figure*}

We consider an uncoded MU-MIMO system used to transmit information streams generated by $N_t$ single-antenna users.  The streams are received by a base station, which is equipped with $N_r\geq N_t$ antennas to simultaneously serve the users. The system is  depicted in  Fig. \ref{up_MU-MIMO}. User $k$ maps $\log_2(\tilde{M})$ bits of its information stream $\qb_k$ to a symbol $\tilde{x}_k \in \tilde{\Omega}$  using a quadrature amplitude modulation (QAM) technique, where $\tilde{\Omega} = \{ s_1, \dots, s_{\tilde{M}}\}$ is a constellation set of $\tilde{M}$-QAM  and $s_m$ is one of the constellation points. The transmitted symbols are uniformly distributed, and the corresponding received signal is given by 
\begin{equation} \label{eII_1a}
\tilde{\qy} = \tilde{\qH} \tilde{\qx} + \tilde{\qn},
\end{equation}
where $\tilde{\qx} = [\tilde{x}_1, \cdots, \tilde{x}_{N_t}]^T$, $\tilde{\qy}=[\tilde{y}_1, \ldots, \tilde{y}_{N_r}]^{T}$,  $\tilde{\qH}=[\tilde{\qh}_1,\dots, \tilde{\qh}_k, \ldots, \tilde{\qh}_{N_t}]  \in \mathbb{C}^{N_r \times {N_t}} $ is the coefficient  matrix of complex memoryless Rayleigh fading channels between ${N_t}$ transmit and $N_r$ receive antennas,  $\tilde{\qh}_k$ is the $k$-th column vector of matrix $\tilde{\qH}$ that denotes wireless channel coefficients between the receive antennas and the $k$-th transmit antenna, where each coefficient follows a Gaussian distribution with zero mean and unity variance, and $\tilde{\qn} \in \mathbb{C}^{N_r}$ denotes the additive white Gaussian noise (AWGN) with a zero mean and covariance matrix $\tilde{\sigma}^2 \qI_{N_r}$. 
The SNR of the system is defined as SNR $ = 10 {\sf{log}}_{10} \left( \frac{{N_t} \mathit{\tilde{E}}_s}{\tilde{\sigma}^2} \right) $ dB, where $\mathit{\tilde{E}}_s$ is the energy per transmit antenna. We normalize the total transmit energy so that $N_t \mathit{\tilde{E}}_s =1 $.
 For convenience, the complex-valued variables are transformed into real-valued variables. Accordingly, we define $\qx = [\mathcal{R}(\tilde{\qx})^T \quad \mathcal{I}(\tilde{\qx})^T]^T  \in \mathbb{R}^K $, $\qy = [\mathcal{R}(\tilde{\qy})^T \quad \mathcal{I}(\tilde{\qy})^T]^T \in \mathbb{R}^N$, $\qn = [\mathcal{R}(\tilde{\qn})^T \quad \mathcal{I}(\tilde{\qn})^T]^T \in \mathbb{R}^N$, and 
$\qH=\begin{bmatrix}
  \mathcal{R}(\tilde{\qH}) & -\mathcal{I}(\tilde{\qH}) \\ 
  \mathcal{I}(\tilde{\qH})  & \mathcal{R}(\tilde{\qH}) 
\end{bmatrix} \in \mathbb{R}^{N \times K}$,
where $K = 2 N_t$,  $N=2N_r$, $\mathcal{R}(\cdot)$ and $\mathcal{I}(\cdot)$ are the real and imaginary parts, respectively.
Therefore, we can rewrite \eqref{eII_1a} as 
\begin{equation} \label{eII_1}
\qy = \qH \qx + \qn.
\end{equation}
Note that the covariance matrix of $\qn$ is  $\sigma^2 \qI_{N} \triangleq    (\tilde{\sigma}^2/2)  \qI_{N}$,  the energy per transmit antenna in the real-valued system is $\mathit{E}_s \triangleq \tilde{\mathit{E}_s}/2$, and the real-valued constellation is  $\Omega = \{ \mathcal{R}(s_m) | s_m\in \tilde{\Omega} \}$ with $| \Omega  |= M \triangleq  \sqrt{\tilde{M}}$. 
We consider the system model  \eqref{eII_1} for the rest of the paper.


\section{The Graph Expectation Propagation Network}

In this section, we propose the GEPNet detector  integrating the  EP  \cite{Jespedes-TCOM14} and GNN  \cite{AScotti_GNN_2020} schemes.
As shown in Fig. \ref{F2}, the GEPNet detector consists of  the observation, GNN and estimation modules, which iteratively exchange the outputs (see Fig. \ref{F2}a). 

\subsection{The Observation Module}

The posterior probability distribution of the transmitted symbols conditioned on the received signal in \eqref{eII_1} can be expressed as
\begin{flalign} \label{eq_EP:Posterior_ori}
    & p(\qx|\qy) = \frac{p(\qy|\qx) }{ p(\qy)} \cdot p(\qx) \propto \underbrace{\mathcal{N} \left( \qy: \qH \qx ,  \sigma^2 \qI_{N_r} \right) }_{p(\qy|\qx)}    \underbrace{ \prod_{k=1}^{K} p(x_{k}) }_{p(\qx)},
\end{flalign}
where  $p(x_k) = \frac{1}{M} \sum_{x \in \Omega } \delta(x_k-x)$ is a priori probability density function  of $x_k$, $\delta$ is the Dirac delta function, and $p(\qy)$ is omitted as it is not related to random variable $x_k$.   A direct calculation of \eqref{eq_EP:Posterior_ori} results in an exponential complexity, which is prohibitive. Therefore, the EP scheme is used to approximate $p(\qx|\qy)$ at the $t$-th iteration by a Gaussian posterior function  
\begin{flalign}\label{eq_EP:Post_approx}
p^{(t)}(\qx|\qy) \propto & p(\qy | \qx) \cdot \chi^{(t)}(\qx)\notag \\
 \propto & \mathcal{N}  \left( \qx: {\qH}^\dagger \qy, \sigma^{2} \left( \qH^T \qH \right)^{-1} \right)  \notag \\
&\cdot \mathcal{N}  \left(\qx: (\qlambda^{(t-1)})^{-1} \qgamma^{(t-1)}, (\qlambda^{(t-1)})^{-1}  \right)\notag \\
\propto &\mathcal{N}  \left( \qx:\qmu^{(t)}, \qSigma^{(t)} \right),
\end{flalign}
where  $\chi^{(t)}(\qx)$ is an approximation of $p(\qx)$ obtained from the exponential family \cite{Jespedes-TCOM14},  $\qlambda^{(t)} $ is a $K \times K$ diagonal matrix with diagonal elements $\lambda_{k}^{(t)}>0 $  and $ \qgamma^{(t)} = [\gamma_{1}^{(t)}, \dots, \gamma_{K}^{(t)}]^T$. Both  $\lambda_{k}^{(t)}$ and  $\gamma_k^{(t)}$  are real  numbers with  $\lambda_{k}^{(0)}=1/\mathit{E}_s $ and $\gamma_k^{(0)}=0$. Note that $p(\qy|\qx)$ in \eqref{eq_EP:Post_approx} is approximated by treating $\qx$ as a random real-valued vector. The product of two Gaussians in \eqref{eq_EP:Post_approx} is computed by using the Gaussian product property\footnote{The product of two Gaussians results in another Gaussian, $\mathcal{N}(\qx:\qa,\qA) \cdot \mathcal{N}(\qx:\qb,\qB)  \propto \mathcal{N} (\qx:(\qA^{-1}+\qB^{-1})^{-1}(\qA^{-1} \qa + \qB^{-1} \qb),(\qA^{-1}+\qB^{-1})^{-1}$.}, given in Appendix A.1 of \cite{Rasmussen-BOOK}. Accordingly, we obtain the variance and mean of $p^{(t)}(\qx|\qy) $ as
\begin{subequations} \label{eA1_a0102}
            \begin{align}
&\qSigma^{(t)} =  {  \left( \sigma^{-2} \qH^T \qH+ \qlambda^{(t-1)} \right)}^{-1}, \label{eA1_a01}\\
& \qmu^{(t)} =\qSigma^{(t)} {\left( \sigma^{-2} \qH^T \qy + \qgamma^{(t-1)}\right)}. \label{eA1_a02}
            \end{align}
        \end{subequations}
We then compute the likelihood function $p^{(t)}(\qy|\qx)$ based on the Gaussian posterior function $p^{(t)}(\qx|\qy) $,
 \begin{flalign}\label{eA1_a0304raw}
p^{(t)}(\qy|\qx) & \triangleq \frac{p^{(t)}(\qx|\qy)}{\chi^{(t)}(\qx)} \notag\\ &
\propto\frac{\mathcal{N}  \left( \qx:\qmu^{(t)}, \qSigma^{(t)} \right)}   {\mathcal{N} \left(\qx:(\qlambda^{(t-1)})^{-1} \qgamma^{(t-1)}, (\qlambda^{(t-1)})^{-1} \right)} 
\notag\\ & \propto \mathcal{N}  \left( \qx: \qx^{(t)}_{{\rm obs}}, \qV^{(t)}_{{\rm obs}} \right),
 \end{flalign}
 where $ \qx^{(t)}_{\rm obs} = [x^{(t)}_{{\rm obs},1}, \dots,x^{(t)}_{{\rm obs},K}] $ and   $\qV^{(t)}_{\rm obs}$ is a $K \times K$ diagonal matrix with  $v^{(t)}_{ {\rm obs},k}$ as the $k$-th diagonal element, which can be expressed as
 \begin{subequations} \label{eA1_a0304}
            \begin{align}
&v_{{\rm obs},k}^{(t)} =  \frac{\Sigma_{k}^{(t)} }{1- \Sigma_{k}^{(t)}  \lambda_{k}^{(t-1)}},  \label{eA1_a03}\\
&x_{{\rm obs},k}^{(t)}  = v_{{\rm obs},k}^{(t)}  {\left(\frac{\mu_{k}^{(t)}}{\Sigma_{k}^{(t)}}-\gamma_{k}^{(t-1)}\right)}.  \label{eA1_a04} 
            \end{align}
        \end{subequations}
Here, $\mu_k^{(t)}$ is the $k$-th element of vector $\qmu^{(t)}$ and $\Sigma_k^{(t)}$ is the $k$-th diagonal element of matrix $\qSigma^{(t)}$.
We treat the pair $\left(\qx_{{\rm obs}}^{(t)},\qV_{{\rm obs}}^{(t)}\right)$ from \eqref{eA1_a0304} as a prior information for the variable nodes $x_1,\dots, x_K$ in the GNN module, as shown in Fig \ref{F2}b.
 
 \subsection{The GNN Module}

 The GNN module employs the message passing (MP) scheme of the pair-wise MRF model \cite{AScotti_GNN_2020}, as described in the Fig. \ref{F2}b. The variable and factor nodes of the GNN are displayed as circles and rectangles, respectively. As in a pair-wise MRF, the $k$-th variable node is characterized by a self potential $\phi(x_k)$, and the $(k,j)$-th pair of variable nodes is characterized by a pair potential $\psi(x_k,x_j)$, where
 \begin{subequations}\label{phi_and_psi}
            \begin{equation}\label{phi}
\phi(x_k) ={ \sf exp} \left( \frac{1}{\sigma^2} \qy^T \qh_k x_k - \frac{1}{2}  \qh_k^T \qh_k x_k^2  \right) p(x_k),
            \end{equation}
            \begin{equation}\label{psi}
\psi(x_k,x_j) = {\sf exp} \left(- \frac{1}{\sigma^2} \qh_k^T \qh_j x_k x_j\right).
            \end{equation}
\end {subequations}
 The GNN is used to infer the posterior probability of the transmitted symbols by using the mean $x_{{\rm obs,k}}^{(t)}$ and variance $v_{{\rm obs,k}}^{(t)}$ for the Gaussian approximation of $x_k$  obtained from the observation module, $k= 1,\dots,K$.
 The mean and variance  are concatenated as
 \begin{equation}\label{concat}
 \qa_k^{(t)} = \left[ x_{\rm obs,k}^{(t)}, v_{\rm obs,k}^{(t)} \right],
 \end{equation}
 and then $\qa_k^{(t)}$ is added as an attribute to the corresponding variable node $x_k$.
The posterior probability corresponding to the pair-wise MRF can be written as \cite{AScotti_GNN_2020}
\begin{equation}\label{p_x_y_GNN}
 p_{\rm GNN}(\qx|\qy) 
 =\frac{1}{Z} \prod_{k=1}^K \phi(x_k) 
 \prod_{\substack{j=1\\j \neq k}}^K \psi(x_k,x_j), 
\end{equation} 
 where $Z$ is a normalization constant.
To compute $p_{\rm GNN}(\qx|\qy)$ in \eqref{p_x_y_GNN}, we use variable and factor feature vectors corresponding to self and pair potentials in \eqref{phi} and \eqref{psi}, respectively. The variable feature vector is denoted as $\qu_k^{(\ell)}$. Its initial value is obtained from encoding the information of the received signal, corresponding channel vector, and noise variance according to \eqref{phi} using a single layer NN as
 \begin{equation}\label{GNN_init}
 \qu_k^{(0)} = \qW_1 \cdot [\qy^T \qh_k, \qh_k^T \qh_k, \sigma^2]^T + \qb_1,
 \end{equation}
 where $\qW_1 \in \mathbb{R}^{N_u \times 3}$ is a learnable matrix,   $\qb_1 \in \mathbb{R}^{N_u }$ is a learnable vector, and $N_u$ is the size of the feature vector. We consider $N_u = 8$. 
The factor feature vector  $\mathbf{f}_{jk} \triangleq \left[ \qh_k^T \qh_j, \sigma^2 \right] $ is obtained by extracting the pair potential information from   \eqref{psi}. The factor feature vector is used  in the MPs of the GNN.  
 As described in Fig. \ref{F2}b, the initialized feature vectors are sent to the corresponding factor nodes. The factor nodes then commence the following iterative MP between the factor and variable nodes:
 
 \subsubsection{\textbf{Factor to variable}}
 
 Each factor node has a multi-layer perceptron (MLP) with two hidden layers of sizes $N_{h_1}$ and $N_{h_2}$ and an output layer of size $N_u$.  In this work, we set  $N_{h_1}=64$ and $N_{h_2}=32$. The rectifier linear unit (ReLU) activation function is used at the output of each hidden layer. For any pair of variable nodes $x_k$ and $x_j$, there is a factor node connecting them. This factor node first concatenates the received feature vectors $\qu_k^{(\ell-1)}$ and $\qu_j^{(\ell-1)}$ with its own feature vector  $\mathbf{f}_{jk}$.
 The factor node then uses the concatenated features as inputs for its MLP, denoted as $\sf{D}$, and saves the corresponding output, expressed as 
 \begin{equation}\label{factor_to_var}
 \qm_{jk}^{(\ell)} = {\sf{D}} \left( \qu_k^{{(}\ell-1{)}}, \qu_j^{{(}\ell-1{)}}, \mathbf{f}_{jk} \right).
 \end{equation}
Finally, the outputs are fed back to the variable nodes as illustrated in Fig. \ref{F2}b. 
 
 \subsubsection{\textbf{Variable to factor}}
 
 The $k$-th variable node then sums all the incoming messages  from its neighbouring factor nodes $\qm_{jk}^{(\ell)}$ and concatenates their sum with the node attribute $\qa_k^{(t)}$ as $\qm_k^{(\ell)}  = \left[\sum_{\substack{j=1\\j \neq k}}^K \qm_{jk}^{(\ell)},  \qa_k^{(t)} \right]$. The concatenated vector is  used to compute the node feature vector $\qu_k^{(\ell)}$ as
 \begin{subequations}\label{GRU_MLP}
            \begin{equation}\label{GRU}
 \qg_k^{(\ell)} = {\sf{U}} \left( \qg_k^{(\ell-1)}, \qm_k^{(\ell)}  \right)
            \end{equation}
            \begin{equation}\label{MLP}
 \qu_k^{(\ell)}= \qW_2 \cdot \qg_k^{(\ell)} + \qb_2,
            \end{equation}
\end {subequations}
where function $\sf U$ is specified by the gated recurrent unit (GRU) network \cite{KYoon2019}, whose current and previous hidden states are $\qg_k^{(\ell)} \in \mathbb{R}^{N_{h_1} }$ and  $\qg_k^{({\ell}-1)} \in \mathbb{R}^{N_{h_1} }$, respectively, $\qW_2 \in \mathbb{R}^{N_u \times N_{h_1}}$ is a learnable matrix, and  $\qb_2\in \mathbb{R}^{N_u }$ is a learnable vector. The updated feature vector  \eqref{MLP} is  then sent to the neighbouring factor nodes to continue the  MP iterations. 

After $L$ rounds of the MP, a readout process yields 
 \begin{subequations}\label{Readout}
				\begin{equation}\label{Readout1}
\tilde p_{\rm GNN}(x_k=a|\qy)  = {\sf{R}} \left(  \qu_k^{(L)} \right), a\in \Omega,
				\end{equation}
            	\begin{equation}\label{Readout2}
p_{\rm GNN}^{(t)}(x_k=a|\qy)  = \frac{{\sf{exp}} \left(\tilde p_{\rm GNN}(x_k=a|\qy) \right)}  {\sum_{b\in \Omega}  {\sf{exp}} \left({\tilde p}_{\rm GNN}(x_k=b|\qy) \right)}, a\in \Omega.
           		 \end{equation}
\end {subequations}
In this work, we set $L=2$.
The readout function ${\sf{R}} $ is given by an MLP with two hidden layers of sizes $N_{h_1}$ and $N_{h_2}$, and ReLU activation at the output of each hidden layer. The output size of  ${\sf{R}} $ is the cardinality of real-valued constellation set, i.e., $M$.
We then assign
\begin{equation}\label{GNN_reset_index}
 \qg_k^{(0)}  \leftarrow   \qg_k^{(L)} \text { and }\qu_k^{(0)} \leftarrow   \qu_k^{(L)}, k= 1, \dots,K
\end{equation}
in order to use the GRU hidden state  and variable feature vector as the starting point for the next GEPNet iteration.
 \begin{Remark}
 In the EP detector, the posterior probability distribution of the transmitted symbol  estimates is calculated as $p(\qx|\qy)\propto \prod_{k=1}^K f_{[x_{{\rm obs},k}^{(t)},v_{{\rm obs},k}^{(t)}]} \left(x_k  \right)$, where $f_{[x_{{\rm obs},k}^{(t)},v_{{\rm obs},k}^{(t)}]}(\cdot)$ is a Gaussian function parameterized by mean $x_{{\rm obs},k}^{(t)}$ and variance $v_{{\rm obs},k}^{(t)}$. 
 In our proposed detector, we replace the Gaussian function with a GNN function parameterized not only by $x_{{\rm obs},k}^{(t)},v_{{\rm obs},k}^{(t)}$, but also by $\qy^T\qh_k,\qh_k^T\qh_k,\qh_k^T\qh_j,\sigma^2$, where $j=1, \dots, K$, $j \neq k$. 
 The GNN function gives more diversity when calculating the posterior probability distribution of the transmitted symbol estimates and enables the proposed detector to capture the MUI information characterized by the pair potential feature $\qh_k^T\qh_j$. 
 \end{Remark}
 
\subsection{The Estimation Module} 

The soft symbol estimate and its variance are computed as  \cite{LeonGarcia-BOOK}
 \begin{subequations}\label{eA1_b0102}
            \begin{equation}
\hat{x}_k^{(t)}=   \sum_{a\in \Omega}  a \times p_{\rm GNN}^{(t)}(x_k=a|\qy),
            \end{equation}
            \begin{equation}
v_k^{(t)} =  \Ex  \left[ \left| x_k  -\hat{x}_k^{(t)}\right|^{2} \right],
            \end{equation}
\end {subequations}
for each $1\leq k\leq K$. 
 We define a vector  $ \hat{\qx}^{(t)}= [\hat{x}_1^{(t)}, \dots,\hat{x}_K^{(t)}] $ and a $K \times K$ diagonal matrix  $\qV^{(t)}$  with  $v_k^{(t)} $ in the $k$-th diagonal element, $k= 1,\dots, K$. 
The work of the GEPNet detector is finished once the maximum number of iterations  $T$ has been reached. Hard estimates of the  transmitted symbols are then made  from $\hat{\qx}^{(T)}$ by comparing   their Euclidean distance from the symbol set $\Omega$.

In the case of $t \neq T$, the Gaussian posterior function ${p}^{(t)}(\qx|\qy)$ is re-evaluated  by updating $\chi^{(t)}(\qx)$  as \cite{Jespedes-TCOM14}
 \begin{flalign} \label{eq:EP_inference_recon}
\chi^{(t+1)}(\qx)
 &\propto  \frac{\mathcal{N}  \left( \qx:  \hat{\qx}^{(t)}, \qV^{(t)} \right)}{\mathcal{N} \left( \qx: \qx^{(t)}_{{\rm obs}},\qV^{(t)}_{{\rm obs}}\right) } \notag \\
 &=  \mathcal{N}  \left(  \qx: (\qlambda^{(t)})^{-1} \qgamma^{(t)},  (\qlambda^{(t)})^{-1}   \right),
\end{flalign} 
where the parameters
\begin{subequations} \label{eA1_b0304}
            \begin{align}
&\qlambda^{(t)} = (\qV^{(t)})^{-1} -  (\qV^{(t)}_{{\rm obs}})^{-1},  \label{eA1_b03}\\
&\qgamma^{(t)} =  (\qV^{(t)})^{-1} \hat{\qx}^{(t)} - (\qV^{(t)}_{{\rm obs}})^{-1}   \qx^{(t)}_{{\rm obs}}. \label{eA1_b04}
            \end{align}
\end{subequations}
Note that $\qlambda^{(t)}$ in \eqref{eA1_b03} may yield a negative value, which should not be the case as it is inverse variance term \cite{Jespedes-TCOM14}. Therefore, when $\lambda_k^{(t)} <0$, we assign $\lambda_k^{(t)}=\lambda_k^{(t-1)}$ and $\gamma_k^{(t)}=\gamma_k^{(t-1)}$. Finally, we smoothen the update of $(\qlambda^{(t)},\qgamma^{(t)})$ by using a convex combination with the former values,
\begin{subequations} \label{eq:damping}
\begin{align}
   \qlambda^{(t)} &= (1-\eta)\qlambda^{(t)}+\eta \qlambda^{(t-1)}, \label{damping_lambda} \\
   \qgamma^{(t)} &= (1-\eta)\qgamma^{(t)}+\eta\qgamma^{(t-1)}\label{damping_gamma},
\end{align}
\end{subequations}
where $\eta\in[0, 1]$ is a weighting coefficient.
The estimation module sends the parameters $(\qgamma^{(t)},\qlambda^{(t)})$ to the observation module, as illustrated in Fig. \ref{F2}a. The complete pseudo-code is shown in Alg. \ref{A1}. 

\begin{algorithm}
\caption{GEPNet detector }
\label{A1}
\begin{algorithmic}[1]
\State {\textbf{Input: } $\qH,\qy,\sigma^2,\mathit{E}_s ,L,T$} 
\State {Initialization:  $\qgamma^{(0)} = \qzero ,  \qlambda^{(0)} =\frac{1}{\mathit{E}_s }\textbf{I} , \eta = 0.7,\qg_k^{(0)}=\bold{0}$}
	\For {$t=1,\dots,T$} 
	
	    \Statex \textbf{\quad\, The Observation Module:}
		\State{Compute  $\qSigma^{(t)}$ and $\qmu^{(t)}$  in \eqref{eA1_a0102}}
		\State{Compute  $v_{{\rm obs},k}^{(t)}$  and $x_{{\rm obs},k}^{(t)}, k=1,\dots,K$ in \eqref{eA1_a0304}}

		\Statex \textbf{\quad\, The GNN Module:}
		\State{Compute \eqref{concat}}
		\If {$t=1$}
				\State{Compute $\qu_k^{(0)}, k=1,\dots,K,$ in \eqref{GNN_init}}
		\EndIf
		\For {$l=1,\dots,L$}
				\State{Compute $ \qm_{jk}^{(\ell)}$ in \eqref{factor_to_var}}, $ j,k=1,\dots,K,j\neq k$
				\State{Compute $\qg_k^{(\ell)} $ and  $\qu_k^{(\ell)}$  in \eqref{GRU_MLP}}, $ k=1,\dots,K$
		\EndFor
		
		\State{Compute $p_{\rm GNN}^{(t)}(x_k|\qy)$  in \eqref{Readout}}, $ k=1,\dots,K$
		\Statex \textbf{\quad\, The Estimation Module:}
		\State{Compute $v_k^{(t)}$ and $\hat{x}_k^{(t)}$   in \eqref{eA1_b0102}}, $ k=1,\dots,K$
		\State{Compute \eqref{GNN_reset_index}}	
		\State{Compute $ \qlambda^{(t)} $ and $\qgamma^{(t)}$   in \eqref{eA1_b0304}}
		\If {$\lambda^{(t)}_k <0 $}
			\State{ $\lambda_k^{(t)}=\lambda_k^{(t-1)}$ and $\gamma_k^{(t)}=\gamma_k^{(t-1)}, k=1,\dots,K$}
		\EndIf
        \State{Smoothen $\qlambda^{(t)}$ and $\qgamma^{(t)}$ using \eqref{eq:damping}}
	\EndFor
\State {\textbf{Return: }  Hard symbol estimates from $\left[\hat{x}_1^{(T)}, \dots,\hat{x}_K^{(T)} \right]$}
\end{algorithmic}
\end{algorithm}

\section{Computational Complexity Analysis}
\label{sComplAnalysis}

\begin{table}\small
\centering
\label{table1}
\begin{tabular}{|l|c|} \hline 
 Detector & Complexity  \\ \hline \hline 
AMP \cite{2009Donoho_ProcSci_AMP} & $\mathcal{O} (NKT)$  \\ \hline
GNN \cite{AScotti_GNN_2020} & $\mathcal{O} ((N+S_u N_{h_1} + N_{h_1} N_{h_2} + N_{h_2} S_u)KT)  $ \\ \hline
MMSE \cite{LMMSE} & $\mathcal{O} (K^3+NK^2)$   \\ \hline
RE-MIMO \cite{2021_KPratik_TSP_REMimo} & $\mathcal{O} ((N^2K+NK^2)T)$   \\ \hline
OAMP-Net \cite{2018HHE_Globecom_OAMPNet}  & $\mathcal{O} ((N^3 + K^3 + N K^2 +N^2K) T)$  \\ \hline
EP  \cite{Jespedes-TCOM14} & $\mathcal{O} ((K^3 +NK^2 + M K )T)$    \\ \hline
GEPNet   & \multicolumn{1}{m{6cm}|}{$\mathcal{O} \Big(\big( K^3+NK^2 + M K +(N+S_u N_{h_1} + N_{h_1} N_{h_2} + N_{h_2} S_u )K L \big)T \Big)$ } \\ \hline
ML \cite{ML} & $\mathcal{O} (M^K)$ \\ \hline
 \end{tabular}
\caption{The computational complexity comparison}
\end{table}

In this section, we analyse the computational complexity of the proposed GEPNet detector depicted in Alg. \ref{A1}. Note that we provide complexity for the real-valued system \eqref{eII_1}. The corresponding complexity for the complex-valued system \eqref{eII_1a} can be easily obtained by substituting $K=2N_t$ and $N=2N_r$.  The dominant complexity of the GEPNet detector per iteration is $\mathcal{O} (NK^2 +K^3)$, which comes from \eqref{eA1_a01}. Expressions \eqref{eA1_a02}, \eqref{eA1_b0102}, \eqref{eA1_b0304}, and \eqref{eq:damping} are all related to matrix-vector multiplications and the cost is $\mathcal{O} (K^2+NK+M K)$.  The rest of the operations belong to the GNN computations, whose complexity is $\mathcal{O} ((N+S_u N_{h_1} + N_{h_1} N_{h_2} + N_{h_2} S_u)KL)$. As \eqref{eA1_a0102}-\eqref{eq:damping} are performed  $T$ times, the total computational complexity of the GEPNet detector is   $\mathcal{O} \Big(\big( NK^2 +K^3+M K  + (N+S_u N_{h_1} + N_{h_1} N_{h_2} + N_{h_2} S_u )K L \big)T \Big)$. Table I shows the computational complexity of the proposed detector  in comparison with the state-of-the-art detectors.

\section{Simulation Results}

\begin{figure}
\centering
\subfloat[$N=32, \text{and } K=32$]
{\includegraphics[scale=0.44]{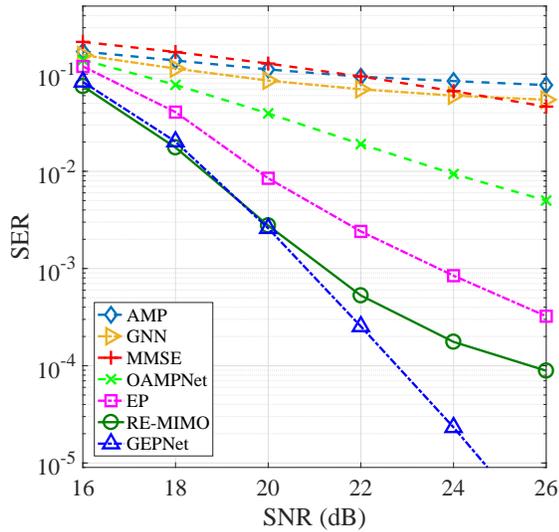}}\hfill
\centering
\subfloat[$N=64, \text{and } K=64$]
{\includegraphics[scale=0.44]{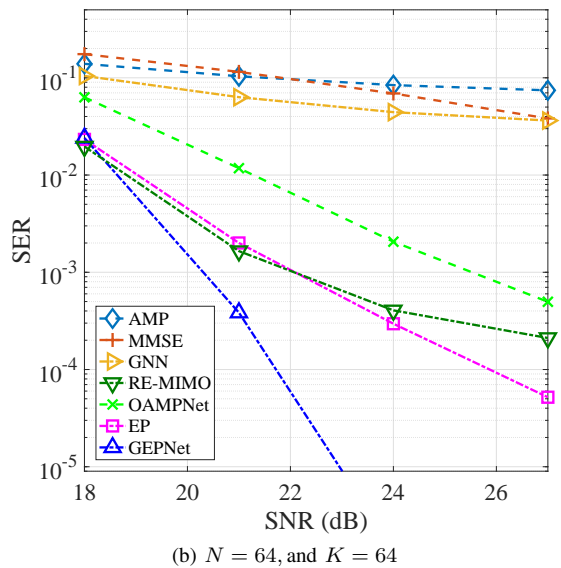}}
\caption{The SER performance comparison}
\label{Results}
\end{figure}

In this section, we explain the training and testing of the NN-based detectors and compare the performance of our proposed detector with the other MU-MIMO detectors.

\subsection{Implementation Details}
We implemented the NN-based detectors OAMPNet, RE-MIMO, GNN, and GEPNet  in PyTorch \cite{Pytorch}. The hyper-parameters for  the existing NN-based detectors were set as in their respective papers. The number of realizations/samples in the training dataset was $80000$ for all the NN-based detectors. The samples were obtained by using QAM modulation with varying SNR values. We applied Adam optimizer with learning rate $0.0001$ to train the proposed detector, and used the total cross-entropy loss function expressed as 
\begin{equation}\label{Loss_func}
Loss= -\frac{1}{Q}  \sum_{q=1}^Q \sum_{k=1}^K  \sum_{a \in \Omega}  \mathbb{I}_{x_k^{(q)}=a}
{\sf log} \left( p_{\rm GNN}^{(T)} \left(x_k=a |\qy^{(q)} \right) \right),
\end{equation}
where $Q$ is the number of training samples in each batch,  $\mathbb{I}_{x_k^{(q)}=a}$ is the indicator function that takes value one
if $x_k^{(q)}=a$ and zero otherwise,  $\qx^{(q)}\in \Omega^K$ is the transmitted vector, $\qy^{(q)}$ is the received signal, and $ p_{\rm GNN}^{(T)} \left(x_k=a |\qy^{(q)} \right) $ is the corresponding probability estimate obtained by the GEPNet detector for the ${q}$-th training sample and $k$-th user. Note that $\mathbb{I}_{x_k^{(q)}=a}$ is used as a training label. 
The GEPNet was trained by using mini-batches of $64$ samples and validated by using $20000$ samples in every epoch.  The total number of epochs was $700$. In the testing phase, we first created a testing dataset by randomly generating $1000000$ samples for the same system configurations ($K,N,M$) that were used in the training phase for each SNR point. Finally, we tested  all the trained detectors using the testing dataset.

\subsection{SER Comparisons}

We investigate the SER performance of our proposed detector by comparing it with those of the MMSE \cite{LMMSE}, AMP \cite{2009Donoho_ProcSci_AMP} and EP \cite{Jespedes-TCOM14}, unfolded NN-based OAMPNet \cite{2018HHE_Globecom_OAMPNet}, RE-MIMO \cite{2021_KPratik_TSP_REMimo} detectors.  We use $16$-QAM modulation scheme. In Fig. \ref{Results}, we employ  $N=K=32$ and   $N=K=64$. The AMP, MMSE, and GNN detectors perform poorly under this system configuration, as well as under other configurations with high ratios of transmit-to-receive antennas. 
The classiscal EP detector is able to achieve a better SER performance than the advanced NN-based OAMPNet detector. 
This is because the EP detector has a significantly better performance compared to the classical AMP based detector.
It can be seen from Fig. \ref{Results} that the proposed detector  achieves at least $4$ dB performance gain compared to the EP detector at  SER of $ 10^{-4}$. 
We observe that the curves in Figs. \ref{Results}a-b behave in a similar way. 
From these facts, we conclude that the GEPNet detector has a significant performance improvement over the state-of-the-art MU-MIMO detectors.

\section{Conclusion}

We proposed a high performance MU-MIMO detector, referred to as the GEPNet detector.  Simulation results showed that the SER performance of the GEPNet detector was significantly better than that of  the other MU-MIMO detectors.

\section*{Acknowledgment}

This research was supported by the research training program stipend  from the University of Sydney.  The work of Branka Vucetic was supported by the Australian Research Council Laureate Fellowship grant number FL160100032.

%

%

{\renewcommand{\baselinestretch}{1.1}
\begin{footnotesize}
\bibliographystyle{IEEEtran}
\bibliography{IEEEabrv,myBib}
\end{footnotesize}}

\end{document}